\documentclass[submission,Ken Akiba,creativecommons]{eptcs}
\providecommand{\event}{6th International Workshop on Classical Logic and Computation 2016} % Name of the event you are submitting to
\usepackage{breakurl}             % Not needed if you use pdflatex only.
\usepackage{microtype}%if unwanted, comment out or use option "draft"
\usepackage{bussproofs}
\usepackage{amsmath}
\usepackage{MnSymbol}

\title{Denotational Semantics of the Simplified Lambda-Mu Calculus and a New Deduction System of Classical Type Theory}
\author{Ken Akiba
\institute{Department of Philosophy\\Virginia Commonwealth University\\Richmond, VA 23284-2025, USA}
\email{kakiba@vcu.edu}
}
\def\titlerunning{Denotational Semantics of Lambda-Mu Calculus}
\def\authorrunning{Ken Akiba}
\begin{document}
\maketitle

\begin{abstract}
\textit{Classical} (or \textit{Boolean}) type theory is the type theory that allows the type inference $(\sigma \to \bot )\to \bot \Rightarrow \sigma $ (the type counterpart of double-negation elimination), where $\sigma $ is any type and $\bot $ is absurdity type. This paper first presents a denotational semantics for a simplified version of Parigot's lambda-mu calculus, a premier example of classical type theory. In this semantics the domain of each type is divided into infinitely many \textit{ranks} and contains not only the usual members of the type at rank 0 but also their negative, conjunctive, and disjunctive \textit{shadows} in the higher ranks, which form an infinitely nested Boolean structure.  Absurdity type $\bot $ is identified as the type of truth values.  The paper then presents a new deduction system of classical type theory, a sequent calculus called \textit{the classical type system (CTS)}, which involves the standard logical operators such as negation, conjunction, and disjunction and thus reflects the discussed semantic structure in a more straightforward fashion.
\end{abstract}

\section{Introduction}

\textit{Classical} (or \textit{Boolean}) type theory is the type theory that allows the type inference $(\sigma \to \bot )\to \bot \Rightarrow \sigma $, where $\sigma $ is any type and $\bot $ is absurdity type. It is so called because this inference rule is the type counterpart of the double-negation elimination rule (where $\neg \sigma=\sigma \to \bot$), the signature inference rule of classical logic in contrast to minimal or intuitionistic logic, which is the usual logic of type inferences. The most well known and influential deduction system of classical type theory to date is M. Parigot's \cite{parigot} $\lambda \mu$-calculus. Section 2 of this paper presents a simplified version of the simply-typed $\lambda \mu$-calculus, and Section 3 gives a denotational semantics to this calculus.  In this semantics the domain of each type is divided into infinitely many \textit{ranks} and contains not only the usual members of the type at rank 0 but also their negative, conjunctive, and disjunctive \textit{shadows} in the higher ranks, which form an \textit{infinitely nested Boolean structure}. Absurdity type $\bot $ is identified as the type of truth values (in contrast to the intuitionistic case, in which it is identified as the empty type).  Section 4 then presents a new deduction system of classical type theory, a sequent calculus called \textit{the classical type system (CTS)}, which involves the standard logical operators such as negation, conjunction, and disjunction and thus reflects the semantic structure given to the simplified $\lambda \mu$-calculus in a more straightforward fashion.

To build an infinitely nested Boolean structure in type $\sigma$, we start with the usual members of the type at rank 0; we then add to the domain the Boolean operators ${-}_{\sigma}^{1}$ (complement), ${\sqcap}_{\sigma}^{1}$ (infimum), and ${\sqcup}_{\sigma}^{1}$ (supremum) at rank 1, creating the negative, conjunctive, and disjunctive objects (\textit{shadows}); we then add another set of Boolean operators ${-}_{\sigma}^{2}, {\sqcap}_{\sigma}^{2}$, and ${\sqcup}_{\sigma}^{2}$, creating the further negative, conjunctive, and disjunctive shadows, and so on, \textit{ad infinitum}. Absurdity type $\bot$ (= the type of truth values) has the same structure, except that it has only two members, 0 (or Falsity) and 1 (or Truth), at rank 0. We shall exploit the fact that, thanks to the binary nature of $\bot$, for any type $\sigma$, type $(\sigma \to \bot )\to \bot$, rank $n$, is isomorphic to type $\sigma $, rank $n+1$.

\section{The simplified simply-typed $\lambda \mu$-calculus }

This section describes the simplified version of the simply-typed $\lambda \mu$-calculus ($S \lambda \mu$), to which a semantics will be given in the next section. This version is simplified from Parigot's original simply-typed $\lambda \mu$-calculus, dropping the original distinction between $\lambda$- and $\mu$-variables. Also, Parigot's original formalization of the calculus, especially his formalization of its inference rules, is a little difficult to follow; so we employ a formalization more familiar to many. Henceforth the qualification \textquoteleft simply-typed' will be omitted for the sake of simplicity. We first present the calculus as a Gentzen-type sequent calculus, making clear why it should be considered a calculus of classical type theory; but we then proceed to turn it into a Hilbert-type axiomatic system of equality, to which a semantics can be more easily given.

\subsection{Language}
\begin{itemize}
\item Types. As usual, there are two kinds of types -- basic (or atomic) types, denoted as $\sigma$ below, and function types. The types are defined thus: 
\[{\tau} \ ::= \ {\sigma } \ | \ {\tau \to \tau } \ | \ {\bot}\]
Greek lower-case alphabets and $\to$ are used to name types with the exception of $\bot$. $\bot$ is called \textit{absurdity type}. Throughout this paper, $\sigma \to \bot$ is abbreviated as $\neg \sigma$.
\item Variables. The last part of Roman lower-case alphabets, \textit{x}, \textit{y}, \textit{z}, ..., are used for variables. The variables not bound by $\lambda$ or $\mu$ are free variables ($FV$).
\item Terms. For clarity's sake, we stipulate that all terms are subscripted with their types. The following is the definition of a term ${{P}_{\sigma }}$ (of type $\sigma$):
\[{{P}_{\sigma }} \ ::= \ {{x}_{\sigma }} \ | \ {{({{P}_{\sigma \to \tau }}{{P}_{\sigma }})}_{\tau }} \ | \ {{(\lambda {{x}_{\sigma }}.{{P}_{\tau }})}_{\sigma \to \tau }} \  | \ {{(\mu {{x}_{\neg \sigma }}.{{P}_{\bot }})}_{\sigma }}\]
Here the same Greek alphabets in the same item are the same types. The first and middle parts of Roman upper-case alphabets, \textit{A}, \textit{B}, \textit{C}, ..., \textit{P}, \textit{Q}, \textit{R}, ..., are used to name terms.
\end{itemize}

\subsection{Deduction system}
The original idea. The simplified $\lambda\mu$-calculus, presented here as a sequent calculus, has the following four inference rules -- axiom, $\lambda$-application and -abstraction, and $\mu$-abstraction:\medskip

\begin{center}
\AxiomC{}
\RightLabel{$Ax$}
\UnaryInfC{$\Gamma ,\ {{x}_{\sigma }}\Rightarrow {{x}_{\sigma }}$}
\DisplayProof \medskip
\end{center}
\begin{center}
\AxiomC{$\Gamma \Rightarrow {{P}_{\sigma \to \tau }}$}
\AxiomC{$\Gamma \Rightarrow {{Q}_{\sigma }} $}
\RightLabel{$\lambda Ap$}
\BinaryInfC{$\Gamma \Rightarrow {{({{P}_{\sigma \to \tau }}{{Q}_{\sigma }})}_{\tau }}$}
\DisplayProof
\quad \quad \quad \quad \quad \quad \quad \quad
\AxiomC{$\Gamma , \ {{x}_{\sigma }}\Rightarrow {{P}_{\tau }}$} 
\RightLabel{$\lambda Ab$}
\UnaryInfC{$\Gamma \Rightarrow {{(\lambda {{x}_{\sigma }}.{{P}_{\tau }})}_{\sigma \to \tau }}$}
\DisplayProof
\end{center}

\begin{center}
\AxiomC{$\Gamma , \ {{x}_{\neg \sigma }}\Rightarrow {{P}_{\bot }}$} 
\RightLabel{$\mu Ab$}
\UnaryInfC{$\Gamma \Rightarrow {{(\mu {{x}_{\neg \sigma }}.{{P}_{\bot }})}_{\sigma }}$}
\DisplayProof
\end{center}\medskip

\noindent Here $\Gamma$ is a \textit{set} of typed variables (as opposed to a sequent, thus obviating the structural rules). 

$\lambda$-application and -abstraction are included in the standard $\lambda$-calculus, whereas $\mu$-abstraction is not. The former offer as the type inference rules \textquoteleft if $\Rightarrow \sigma \to \tau$ and $\Rightarrow \sigma$ then $\Rightarrow \tau$' and \textquoteleft if $\sigma \Rightarrow \tau$ then $\Rightarrow \sigma \to \tau$', i.e., the elimination and introduction rules for conditional. $\mu Ab$ offers as the type inference rule double-negation elimination \textquoteleft if $\neg \sigma \Rightarrow \bot$ then $\Rightarrow \sigma$', the signature inference rule in classical logic. 

This simplified version does not include another inference rule, $\mu$-application, included in Parigot's original version:\medskip
\begin{center}
\AxiomC{$\Gamma , \ {{a}_{\neg \sigma }}\Rightarrow {{P}_{\sigma }}$}
\RightLabel{$\mu Ap$}
\UnaryInfC{$\Gamma , \ {{a}_{\neg \sigma }}\Rightarrow {({[a]_{\neg \sigma }{{P}_{\sigma }})}_{\bot }}$}
\DisplayProof
\end{center}\medskip
where $a$ is a variable of a different kind, a $\mu$-variable, and $[ \cdot ]$ is a certain new operation applicable only to $\mu$-variables. $\mu Ap$ offers as a type inference rule \textquoteleft if $\neg \sigma \Rightarrow \sigma$ then $\neg \sigma \Rightarrow \bot$', which is correct in classical logic but is a redundant rule and does not have a counterpart in classical natural deduction. In these respects, it is a little difficult to see that the Curry-Howard correspondence (see, e.g., S\o{}rensen and Urzyczyn \cite{sorensen-urzyczyn}) holds for Parigot's original $\lambda \mu$-calculus (though it does hold in a certain sense). In contrast, the Curry-Howard correspondence clearly holds for the simplified version. The simplified version is attractive in this respect. It is actually more similar to the calculi presented in, e.g., Rehof and S\o{}rensen \cite{rehof} and Baba et al. \cite{baba} than Parigot's original $\lambda \mu$-calculus in that it contains only one kind (the usual kind) of variables.

We have presented above the simplified $\lambda \mu$-calculus as a sequent calculus; however, we shall \textit{not} use it in that form in what follows. We shall instead consider its axiomatic version, the theory of $\beta \eta \mu$-equality, which consists of the following inference rules and notion of deducibility:\medskip
\begin{itemize}
\item Inference rules.\medskip
\begin{enumerate}
\item $\Rightarrow {{P}_{\sigma }}={{P}_{\sigma }}$;\medskip
\item ${{P}_{\sigma }}={{Q}_{\sigma }}\Rightarrow {{Q}_{\sigma }}={{P}_{\sigma }}$; \medskip
\item ${{P}_{\sigma }}={{Q}_{\sigma }},{{Q}_{\sigma }}={{R}_{\sigma }}\Rightarrow {{P}_{\sigma }}={{R}_{\sigma }}$;\medskip
\item ${{P}_{\sigma \to \tau }}={{Q}_{\sigma \to \tau }}\Rightarrow {{({{P}_{\sigma \to \tau }}{{A}_{\sigma }})}_{\tau }}={{({{Q}_{\sigma \to \tau }}{{A}_{\sigma }})}_{\tau }}$;\medskip
\item ${{A}_{\sigma }}={{B}_{\sigma }}\Rightarrow {{({{P}_{\sigma \to \tau }}{{A}_{\sigma }})}_{\tau }}={{({{P}_{\sigma \to \tau }}{{B}_{\sigma }})}_{\tau }}$; \medskip
\item ${{P}_{\tau }}={{Q}_{\tau }}\Rightarrow {{(\lambda {{x}_{\sigma }}.{{P}_{\tau }})}_{\sigma \to \tau }}={{(\lambda {{x}_{\sigma }}.{{Q}_{\tau }})}_{\sigma \to \tau }}$;\medskip
\item ${{P}_{\bot }}={{Q}_{\bot }}\Rightarrow {{(\mu {{x}_{\neg \sigma }}.{{P}_{\bot }})}_{\sigma }}={{(\mu {{x}_{\neg \sigma }}.{{Q}_{\bot }})}_{\sigma }}$;\medskip
\item $\Rightarrow {{{({(\lambda {{x}_{\sigma }}.{{P}_{\tau }})}_{\sigma \to \tau }}{{Q}_{\sigma }})}_{\tau}} = {{P}_{\tau }}[{{x}_{\sigma }}:={{Q}_{\sigma }}]$ \ \ ($\beta$-quality); \medskip
\item $\Rightarrow {(\lambda x_{\sigma}{(P_{\sigma \to \tau}x_{\sigma})}_{\tau})}_{\sigma \to \tau} = P_{\sigma \to \tau}$ \ \ if $x_{\sigma}\notin FV(P_{\sigma \to \tau})$ \ \  ($\eta$-equality);\medskip
\item $\Rightarrow {{({{Q}_{\neg \sigma }}{{(\mu {{x}_{\neg \sigma }}.{{P}_{\bot }})}_{\sigma }})}_{\bot }}={{P}_{\bot }}[{{x}_{\neg \sigma }}:={{Q}_{\neg \sigma }}]$ \ \ ($\beta_{\mu}$-equality);\medskip
\item $\Rightarrow {{(\mu {{x}_{\neg \sigma }}{{({{x}_{\neg \sigma }}{{P}_{\sigma }})}_{\bot }})}_{\sigma }}={{P}_{\sigma }}$ \ \ if $x_{\neg \sigma}\notin FV(P_{\sigma})$ \ \ ($\eta_{\mu}$-equality);\medskip
\item $\Rightarrow {{({{(\mu {{x}_{\neg (\sigma \to \tau )}}.{{P}_{\bot }})}_{\sigma \to \tau }}{{Q}_{\sigma }})}_{\tau }} = {{(\mu {{y}_{\neg \tau }}.{{P}_{\bot }}[{{({{x}_{\neg (\sigma \to \tau )}}{{R}_{\sigma \to \tau }})}_{\bot }}{:=^{*}}{{({{y}_{\neg \tau }}{{({{R}_{\sigma \to \tau }}{{Q}_{\sigma }})}_{\tau }})}_{\bot }}])}_{\tau }}$ \ \ ($\mu$-equality),\medskip

where ${{P}_{\bot }}[{{({{x}_{\neg (\sigma \to \tau )}}{{R}_{\sigma \to \tau }})}_{\bot }}{:=^{*}}{{({{y}_{\neg \tau }}{{({{R}_{\sigma \to \tau }}{{Q}_{\sigma }})}_{\tau }})}_{\bot }}]$ is obtained from ${{P}_{\bot }}$ by replacing inductively each subterm of the form ${{({{x}_{\neg (\sigma \to \tau )}}{{R}_{\sigma \to \tau }})}_{\bot }}$ in ${{P}_{\bot }}$ with ${{({{y}_{\neg \tau }}{{({{R}_{\sigma \to \tau }}{{Q}_{\sigma }})}_{\tau }})}_{\bot }}$.
\end{enumerate}\medskip
\item Deducibility. $\Gamma \ {\vdash_{S \lambda\mu}} \ N$, where $\Gamma$ is a set of equations and $N$ is an equation, if and only if $N$ is derivable from $\Gamma$ with the above inference rules.
\end{itemize}\medskip
\noindent This formalization is in fact closer to -- a simplified simply-typed version of -- Parigot's \cite{parigot} untyped (or \textquoteleft pure') $\lambda \mu$-calculus. Among the above rules, those except 7, 10, 11, and 12 consist of the usual $\lambda$-theory of $\beta \eta$-equality.  From the viewpoint of Curry-Howard correspondence, just as $\beta$- and $\eta$-equality are seen as the normalization procedures for minimal and intuitionistic logics, $\mu$-equality can be seen as the additional normalization procedure for classical logic: the application of the double-negation rule to the conditional $\sigma \to \tau$ is reduced to the application of the rule to its component $\tau$ (see, e.g., Prawitz  \cite{prawitz}, pp. 39--40; Troelstra and Schwichtenberg \cite{ts}, p. 183). $\beta_{\mu}$- and $\eta_{\mu}$-equality represent the double-negation versions of the normalization procedures represented by $\beta$- and $\eta$-equality. These rules are based on the symmetry between $(\lambda {{x}_{\neg \sigma }}.{P}_{\bot })_{\neg \neg \sigma }$ and $(\mu {{x}_{\neg \sigma }}.{{P}_{\bot }})_{\sigma }$ with respect to the terms of type $\neg \sigma$.

\section{Denotational semantics of the simplified $\lambda \mu$-calculus}

A model $\mathcal{M}$ of the simplified $\lambda \mu$-calculus is determined by its domain $D$ and interpretation $[\![\cdot ]\!]^{\mathcal{M}}$, i.e., ${\mathcal{M}}=\langle D, \ [\![\cdot ]\!]^{\mathcal{M}}\rangle$.

\subsection{Domains}
The entire domain $D$ of a model is the sum of all domains ${{D}_{\sigma }}$ of all types $\sigma$. For any type $\sigma$, its domain $D_{\sigma }$ is divided into infinitely many \textit{ranks}, indicated by superscripts. The domain of any rank of any type is a superset of the domains of the lower ranks of the same type, and the domain of rank $n$, where $n$ is a non-limit ordinal, forms a Boolean algebra. The rank 0 domain of each type should be considered the usual domain of the type for the simply-typed $\lambda$-calculus; such a domain is infinitely expanded for the simplified $\lambda \mu$-calculus. We call the resulting structure \textit{an infinitely nested Boolean structure} or \textit{an infinite Boolean expansion} of the rank 0 domain. A more specific description of the domains will follow. In the rest of this paper, an object $a \in D_{\sigma}^{n}$, is indicated as $a_{\sigma }^{n}$. Note that any object of some rank of some type also belongs to any domain of a higher rank of the same type; so $a_{\sigma }^{n}$ may belong not to rank $n$ but to a lower rank. Generally $(p_{\sigma \to \tau }^{n}a_{\sigma }^{n})_{\tau }^{n}$ is the value, of type $\tau$, rank $\leq n$, of the function $p_{\sigma \to \tau }^{n}$ applied to the argument $a_{\sigma }^{n}$. In contrast, the superscripts attached to the Boolean operators such as $-$, $\bigsqcap$, and $\bigsqcup$ indicate the types the operators properly belong to.

\begin{itemize}
\item $D_{\bot }^{0}=\{0,1\}$. 
\item $D_{\bot }^{n}$ (where $n$ is a non-limit ordinal) = the smallest superset of $D_{\bot }^{n-1}$ closed under the Boolean operations $-_{\bot }^{n}$ (complement), ${\bigsqcap}{_{\bot}^{n}}$ (infimum), and ${{\bigsqcup} _{\bot }^{n}}$ (supremum) (the two-member cases of which are ${{\sqcap }_{\bot}^{n}}$ and ${{\sqcup}_{\bot}^{n}}$).
\item $D_{\sigma }^{0}=$ the set of (usual) individuals of type $\sigma $ (so $D_{\sigma \to \tau }^{0}$ is the set of functions from $D_{\sigma }^{0}$ to $D_{\tau }^{0}$).
\item $D_{\sigma }^{n}$ (where $n$ is a non-limit ordinal) = the smallest superset of $D_{\sigma }^{n-1}$ closed under the Boolean operations $-_{\sigma }^{n}$, ${\bigsqcap}{_{\sigma}^{n}}$, ${{\bigsqcup} _{\sigma }^{n}}$, ${{\sqcap }_{\sigma}^{n}}$, and ${{\sqcup}_{\sigma}^{n}}$ that satisfy the following conditions: for any type $\tau$ and any objects $a_{\sigma }^{n},b_{\sigma }^{n},p_{\sigma \to \tau }^{n},q_{\sigma \to \tau }^{n}$, and $r_{\sigma \to \tau}^{n-1}$, 
\[\begin{array}{rcl}
 (r_{\sigma \to \tau }^{n-1}(-_{\sigma }^{n}a_{\sigma }^{n})_{\sigma }^{n})_{\tau }^{n} & = & (-_{\tau }^{n}(r_{\sigma \to \tau }^{n-1}a_{\sigma }^{n})_{\tau }^{n})_{\tau }^{n}; \\ 
 (r_{\sigma \to \tau }^{n-1}(a_{\sigma }^{n}\sqcap _{\sigma }^{n}b_{\sigma }^{n})_{\sigma }^{n})_{\tau }^{n} & = & ((r_{\sigma \to \tau }^{n-1}a_{\sigma }^{n})_{\tau }^{n}\sqcap _{\tau }^{n}(r_{\sigma \to \tau }^{n-1}b_{\sigma }^{n})_{\tau }^{n})_{\tau }^{n}; \\ 
 (r_{\sigma \to \tau }^{n-1}(a_{\sigma }^{n}\sqcup _{\sigma }^{n}b_{\sigma }^{n})_{\sigma }^{n})_{\tau }^{n} & = & ((r_{\sigma \to \tau }^{n-1}a_{\sigma }^{n})_{\tau }^{n}\sqcup _{\tau }^{n}(r_{\sigma \to \tau }^{n-1}b_{\sigma }^{n})_{\tau }^{n})_{\tau }^{n}; \\ 
 (r_{\sigma \to \tau }^{n-1}(\bigsqcap\limits_{a}{_{\sigma }^{n}}a_{\sigma }^{n})_{\sigma}^{n})_{\tau }^{n} & = & (\bigsqcap\limits_{a}{_{\tau }^{n}}(r_{\sigma \to \tau }^{n-1}a_{\sigma }^{n})_{\tau }^{n})_{\tau }^{n}; \\ 
 (r_{\sigma \to \tau }^{n-1}(\bigsqcup\limits_{a}{_{\sigma }^{n}}a_{\sigma }^{n})_{\sigma }^{n})_{\tau }^{n} & = & (\bigsqcup\limits_{a}{_{\tau }^{n}}(r_{\sigma \to \tau }^{n-1}a_{\sigma }^{n})_{\tau }^{n})_{\tau }^{n}; \\ 
 ((-_{\sigma \to \tau }^{n}p_{\sigma \to \tau }^{n})_{\sigma \to \tau }^{n}a_{\sigma }^{n})_{\tau }^{n} & = & (-_{\tau }^{n}(p_{\sigma \to \tau }^{n}a_{\sigma }^{n})_{\tau }^{n})_{\tau }^{n}; \\ 
 ((p_{\sigma \to \tau }^{n}\sqcap _{\sigma \to \tau }^{n}q_{\sigma \to \tau }^{n})_{\sigma \to \tau }^{n}a_{\sigma }^{n})_{\tau }^{n} & = & ((p_{\sigma \to \tau }^{n}a_{\sigma }^{n})_{\tau }^{n}\sqcap _{\tau }^{n}(q_{\sigma \to \tau }^{n}a_{\sigma }^{n})_{\tau }^{n})_{\tau }^{n}; \\ 
 ((p_{\sigma \to \tau }^{n}\sqcup _{\sigma \to \tau }^{n}q_{\sigma \to \tau }^{n})_{\sigma \to \tau }^{n}a_{\sigma }^{n})_{\tau }^{n} & = & ((p_{\sigma \to \tau }^{n}a_{\sigma }^{n})_{\tau }^{n}\sqcup _{\tau }^{n}(q_{\sigma \to \tau }^{n}a_{\sigma }^{n})_{\tau }^{n})_{\tau }^{n}; \\ 
 ((\bigsqcap\limits_{p}{_{\sigma \to \tau}^{n}}p_{\sigma \to \tau}^{n})_{\sigma \to \tau }^{n}a_{\sigma }^{n})_{\tau }^{n} & = & (\bigsqcap\limits_{p}{_{\tau}^{n}}(p_{\sigma \to \tau }^{n}a_{\sigma }^{n})_{\tau }^{n})_{\tau }^{n}; \\ 
 ((\bigsqcup\limits_{p}{_{\sigma \to \tau}^{n}}p_{\sigma \to \tau }^{n})_{\sigma \to \tau }^{n}a_{\sigma }^{n})_{\tau }^{n} & = & (\bigsqcup\limits_{p}{_{\tau }^{n}}(p_{\sigma \to \tau }^{n}a_{\sigma }^{n})_{\tau }^{n})_{\tau }^{n}. \\ 
\end{array}\]

These will be called \textit{the expansion conditions for} $\sigma$.\medskip

\item $D_{\sigma }^{l}=\bigcup\limits_{n<l}{D_{\sigma }^{n}}$ (where $l$ is an infinite limit ordinal).
\item ${{D}_{\sigma }}=\bigcup\limits_{m}{D_{\sigma }^{m}}$ (where $m$ is any ordinal).
\item $D=\bigcup\limits_{\sigma}{D_{\sigma }}$ (where $\sigma$ is any type).
\end{itemize}

\noindent We call the members of $D_{\sigma}^{n}$, where $n>0$, \textit{shadows} of the members of $D_{\sigma}^{0}$, which we call \textit{individuals}.  $(-_{\sigma }^{n}a_{\sigma }^{n})_{\sigma }^{n}$ may also be called \textit{the negative shadow} of $a_{\sigma }^{n}$, $(a_{\sigma }^{n}\sqcap _{\sigma }^{n}b_{\sigma }^{n})_{\sigma }^{n}$ and $(a_{\sigma }^{n}\sqcup _{\sigma }^{n}b_{\sigma }^{n})_{\sigma }^{n}$ \textit{the conjunctive} and \textit{disjuncitive shadows} of $a_{\sigma }^{n}$ and $b_{\sigma }^{n}$, $(\bigsqcap\limits_{a}{_{\sigma }^{n}}a_{\sigma }^{n})_{\sigma }^{n} $ and $(\bigsqcup\limits_{a} {_{\sigma }^{n}}a_{\sigma }^{n})_{\sigma }^{n}$ \textit{the conjunctive} and \textit{disjunctive shadows} of the set of $a_{\sigma }^{n}$s.  Shadows are pseudo-objects whose logical behavior is determined by the individuals and the operators contained.

Intuitively, the expansion conditions state that $-_{\sigma }^{n}, {\sqcap }_{\sigma}^{n}, {{\sqcup}_{\sigma}^{n}}, {\bigsqcap}_{\sigma}^{n}$, and ${{\bigsqcup} _{\sigma }^{n}}$ in $(-_{\sigma }^{n}a_{\sigma }^{n})_{\sigma }^{n}, (a_{\sigma }^{n}\sqcap _{\sigma }^{n}b_{\sigma }^{n})_{\sigma }^{n},(a_{\sigma }^{n}\sqcup _{\sigma }^{n}b_{\sigma }^{n})_{\sigma }^{n}, (\bigsqcap _{\sigma }^{n}a_{\sigma }^{n})_{\sigma }^{n} $, and $(\bigsqcup _{\sigma }^{n}a_{\sigma }^{n})_{\sigma }^{n}$ distribute over objects $r_{\sigma \to \tau }^{n-1}$ for any $\tau$, but that if $\sigma$ is the function type of form $\phi \to \psi$, then they will distribute also over objects $b_{\phi }^{n}$.  This is consistent because even if $b_{\phi }^{n}$ here contains one of the Boolean operators, it won't distribute over the original objects above because they are of rank $n$ and not $n-1$. In any $(p_{\sigma \to \tau }^{m}a_{\sigma}^{n})$, where $p_{\sigma \to \tau}$ and $a_{\sigma}$, both containing Boolean operators, are properly of type $m$ and $n$ respectively, one side distributes over the other side regardless, but which distributes over which depends on the superscripts (of the operators) $m$ and $n$: if $m\geq n$, then $p_{\sigma \to \tau }^{m}$ distributes over $a_{\sigma}^{n}$; if $m<n$, the other way around.

Note, furthermore, that when the operators are distributed (or squeezed out), their types change (from $\sigma$ or $\sigma \to \tau$ to $\tau$ in the above schemas), but their ranks do not. And the ranks (plus whether the relevant items are in the left (or functor) side or the right (or argument) side of the application when their ranks are identical) are what determines how the operators are distributed, not the structure of the relevant expressions. For instance, compare:\medskip

\begin{itemize}
\item $([(p_{\sigma \to \neg \tau }^{0}\sqcap _{\sigma \to \neg \tau }^{1}q_{\sigma \to \neg \tau }^{0})_{\sigma \to \neg \tau }^{1}(-_{\sigma }^{1}a_{\sigma }^{0})_{\sigma }^{1}]_{\neg \tau }^{1}[r_{\tau }^{0}\sqcup _{\tau }^{2}s_{\tau }^{0}]_{\tau }^{2})_{\bot }^{2}=$ \\ 
 $([(-_{\bot }^{1}((p_{\sigma \to \neg \tau }^{0}a_{\sigma }^{0})_{\neg \tau }^{0}r_{\tau }^{0})_{\bot }^{0})_{\bot }^{1}\sqcap _{\bot }^{1}(-_{\bot }^{1}((q_{\sigma \to \neg \tau }^{0}a_{\sigma }^{0})_{\neg \tau }^{0}r_{\tau }^{0})_{\bot }^{0})_{\bot }^{1}]_{\bot }^{1}\sqcup _{\bot }^{2}[(-_{\bot }^{1}((p_{\sigma \to \neg \tau }^{0}a_{\sigma }^{0})_{\neg \tau }^{0}s_{\tau }^{0})_{\bot }^{0})_{\bot }^{1}\sqcap _{\bot }^{1}\linebreak(-_{\bot }^{1}((q_{\sigma \to \neg \tau }^{0}a_{\sigma }^{0})_{\neg \tau }^{0}s_{\tau }^{0})_{\bot }^{0})_{\bot }^{1}]_{\bot }^{1})_{\bot }^{2}$.\medskip
\item  $([p_{\neg \tau }^{0}\sqcap _{\neg \tau }^{1}q_{\neg \tau }^{0}]_{\neg \tau }^{1}[(-_{\sigma \to \tau }^{1}a_{\sigma \to \tau }^{0})_{\sigma \to \tau }^{1}(r_{\sigma }^{0}\sqcup _{\sigma }^{2}s_{\sigma }^{0})_{\sigma }^{2}]_{\tau }^{2})_{\bot }^{2}=$ \\ 
 $([(-_{\bot }^{1}(p_{\neg \tau }^{0}(a_{\sigma \to \tau }^{0}r_{\sigma }^{0})_{\tau }^{0})_{\bot }^{0})_{\bot }^{1}\sqcap _{\bot }^{1}(-_{\bot }^{1}(q_{\neg \tau }^{0}(a_{\sigma \to \tau }^{0}r_{\sigma }^{0})_{\tau }^{0})_{\bot }^{0})_{\bot }^{1}]_{\bot }^{1}\sqcup _{\bot }^{2}[(-_{\bot }^{1}(p_{\neg \tau }^{0}(a_{\sigma \to \tau }^{0}s_{\sigma }^{0})_{\tau }^{0})_{\bot }^{0})_{\bot }^{1}\sqcap _{\bot }^{1}\linebreak(-_{\bot }^{1}(q_{\neg \tau }^{0}(a_{\sigma \to \tau }^{0}s_{\sigma }^{0})_{\tau }^{0})_{\bot }^{0})_{\bot }^{1}]_{\bot }^{1})_{\bot }^{2}$.\medskip
 \item $([p_{\neg \tau }^{0}\sqcap _{\neg \tau }^{2}q_{\neg \tau }^{0}]_{\neg \tau }^{1}[(-_{\sigma \to \tau }^{1}a_{\sigma \to \tau }^{0})_{\sigma \to \tau }^{1}(r_{\sigma }^{0}\sqcup _{\sigma }^{2}s_{\sigma }^{0})_{\sigma }^{2}]_{\tau }^{2})_{\bot }^{2}=$ \\
 $([(-_{\bot }^{1}(p_{\neg \tau }^{0}(a_{\sigma \to \tau }^{0}r_{\sigma }^{0})_{\tau }^{0})_{\bot }^{0})_{\bot }^{1}\sqcup _{\bot }^{2}(-_{\bot }^{1}(p_{\neg \tau }^{0}(a_{\sigma \to \tau }^{0}s_{\sigma }^{0})_{\tau }^{0})_{\bot }^{0})_{\bot }^{1}]_{\bot }^{2}\sqcap _{\bot }^{2}[(-_{\bot }^{1}(q_{\neg \tau }^{0}(a_{\sigma \to \tau }^{0}r_{\sigma }^{0})_{\tau }^{0})_{\bot }^{0})_{\bot }^{1}\sqcup _{\bot }^{2}\linebreak(-_{\bot }^{1}(q_{\neg \tau }^{0}(a_{\sigma \to \tau }^{0}s_{\sigma }^{0})_{\tau }^{0})_{\bot }^{0})_{\bot }^{1}]_{\bot }^{2})_{\bot }^{2}$.
\end{itemize}\medskip
 The main structures are marked by the brackets $[\cdot]$ here. On the one hand, the structures of the first two original expressions are different, but the structures of the resulting expressions, in which all the operators are squeezed out, are identical because the ranks of the operators involved are identical. The different structures of the original expressions are retained in the atomic structures of the resulting expressions. On the other hand, the structures of the last two original expressions are identical, but the structures of the resulting expressions are different ($\sqcap$ and $\sqcup$ reversed) because the ranks of the original $\sqcap$ are different (1 versus 2).  (That's the only difference between the last two original expressions.) The entire structures of the resulting expressions are determined not by the structures of the original expressions but by the ranks of the operators. To summarize this formally,\medskip
 
\noindent \textbf{Definition 0} \ \begin{itemize}
\item An expression of a member of the domain $D$ is \textit{atomic}: \[ A_{\sigma }^{0}\ ::=\ a_{\sigma }^{0}\ |\ (A_{\sigma \to \tau }^{0}A_{\sigma }^{0})_{\tau }^{0} \]
where $a_{\sigma }^{0}$ is a (proper) name of a member.
\item An expression is \textit{molecular}: \[ M_{\sigma }^{k}\ ::=\ A_{\sigma }^{0}\ |\ (M_{\sigma \to \tau }^{m}M_{\sigma }^{n})_{\tau }^{\max (m,n)}\ |\ (-_{\sigma }^{k}M_{\sigma }^{m})_{\sigma }^{k}\ |\ (M_{\sigma }^{m}\sqcap _{\sigma }^{k}M_{\sigma }^{n})_{\sigma }^{k}\ |\ (M_{\sigma }^{m}\sqcup _{\sigma }^{k}M_{\sigma }^{n})_{\sigma }^{k}\ |\ ({\bigsqcap}{_{\sigma }^{k}}A_{\sigma }^{0})_{\sigma }^{k}\ |\ ({\bigsqcup}{_{\sigma }^{k}}A_{\sigma }^{0})_{\sigma }^{k} \]
where $m, n \leq k \neq 0$.
\item An expression is \textit{canonical}: \[C_{\sigma }^{k}\ ::=\ A_{\sigma }^{0}\ |\ (-_{\sigma }^{k}C_{\sigma }^{m})_{\sigma }^{k}\ |\ (C_{\sigma }^{m}\sqcap _{\sigma }^{k}C_{\sigma }^{n})_{\sigma }^{k}\ |\ (C_{\sigma }^{m}\sqcup _{\sigma }^{k}C_{\sigma }^{n})_{\sigma }^{k}\ |\ ({\bigsqcap}{_{\sigma }^{k}}A_{\sigma }^{0})_{\sigma }^{k}\ |\ ({\bigsqcup}{_{\sigma }^{k}}A_{\sigma }^{0})_{\sigma }^{k}  \]
where $m, n \leq k \neq 0$.
\end{itemize}

\noindent Then \medskip

\noindent \textbf{Proposition 1} Every molecular expression of a member of the domain $D$ has an equivalent (i.e., co-denotational) canonical expression.

More informally: In the reduction of a molecular expression into a canonical expression, the distribution of the operators is determined purely by their ranks; the structure of the molecular expression is retained in the structures of the atomic cores of the canonical expression.\medskip

Another implication of the expansion conditions worth singling out is the case in which $\tau =\bot$:\medskip

\noindent \textbf{Lemma 2} \ 
For any type $\sigma $ and any objects $a_{\sigma }^{n},b_{\sigma }^{n},p_{\neg \sigma }^{n},q_{\neg \sigma }^{n}$, and $r_{\neg \sigma}^{n-1}$, \pagebreak
\[\begin{array}{rcl}
 (r_{\neg \sigma }^{n-1}(-_{\sigma }^{n}a_{\sigma }^{n})_{\sigma }^{n})_{\bot }^{n} & = & (-_{\bot }^{n}(r_{\neg \sigma }^{n-1}a_{\sigma }^{n})_{\bot }^{n})_{\bot }^{n}; \\ 
 (r_{\neg \sigma }^{n-1}(a_{\sigma }^{n}\sqcap _{\sigma }^{n}b_{\sigma }^{n})_{\sigma }^{n})_{\bot }^{n} & = & ((r_{\neg \sigma }^{n-1}a_{\sigma }^{n})_{\bot }^{n}\sqcap _{\bot }^{n}(r_{\neg \sigma }^{n-1}b_{\sigma }^{n})_{\bot }^{n})_{\bot }^{n}; \\ 
 (r_{\neg \sigma }^{n-1}(a_{\sigma }^{n}\sqcup _{\sigma }^{n}b_{\sigma }^{n})_{\sigma }^{n})_{\bot }^{n} & = & ((r_{\neg \sigma}^{n-1}a_{\sigma }^{n})_{\bot }^{n}\sqcup _{\bot }^{n}(r_{\neg \sigma }^{n-1}b_{\sigma }^{n})_{\bot }^{n})_{\bot }^{n}; \\ 
 (r_{\neg \sigma }^{n-1}(\bigsqcap\limits_{a} {_{\sigma }^{n}}a_{\sigma }^{n})_{\sigma }^{n})_{\bot }^{n} & = & (\bigsqcap\limits_{a} {_{\bot }^{n}}(r_{\neg \sigma }^{n-1}a_{\sigma }^{n})_{\bot }^{n})_{\bot }^{n}; \\ 
 (r_{\neg \sigma}^{n-1}(\bigsqcup\limits_{a} {_{\sigma }^{n}}a_{\sigma }^{n})_{\sigma }^{n})_{\bot }^{n} & = & (\bigsqcup\limits_{a} {_{\bot }^{n}}(r_{\neg \sigma}^{n-1}a_{\sigma }^{n})_{\bot }^{n})_{\bot }^{n}; \\ 
 ((-_{\neg \sigma }^{n}p_{\neg \sigma }^{n})_{\neg \sigma }^{n}a_{\sigma }^{n})_{\bot }^{n} & = & (-_{\bot }^{n}(p_{\neg \sigma }^{n}a_{\sigma }^{n})_{\bot }^{n})_{\bot }^{n}; \\ 
 ((p_{\neg \sigma }^{n}\sqcap _{\neg \sigma }^{n}q_{\neg \sigma }^{n})_{\neg \sigma }^{n}a_{\sigma }^{n})_{\bot }^{n} & = & ((p_{\neg \sigma}^{n}a_{\sigma }^{n})_{\bot }^{n}\sqcap _{\bot }^{n}(q_{\neg \sigma }^{n}a_{\sigma }^{n})_{\bot }^{n})_{\bot }^{n}; \\ 
 ((p_{\neg \sigma }^{n}\sqcup _{\neg \sigma }^{n}q_{\neg \sigma }^{n})_{\neg \sigma }^{n}a_{\sigma }^{n})_{\bot }^{n} & = & ((p_{\neg \sigma }^{n}a_{\sigma }^{n})_{\bot }^{n}\sqcup _{\bot }^{n}(q_{\neg \sigma }^{n}a_{\sigma }^{n})_{\bot }^{n})_{\bot }^{n}; \\ 
 ((\bigsqcap\limits_{p} {_{\neg \sigma }^{n}}p_{\neg \sigma }^{n})_{\neg \sigma }^{n}a_{\sigma }^{n})_{\bot }^{n} & = & (\bigsqcap\limits_{p} {_{\bot }^{n}}(p_{\neg \sigma }^{n}a_{\sigma }^{n})_{\bot }^{n})_{\bot }^{n}; \\ 
 ((\bigsqcup\limits_{p} {_{\neg \sigma }^{n}}p_{\neg \sigma }^{n})_{\neg \sigma }^{n}a_{\sigma }^{n})_{\bot }^{n} & = & (\bigsqcup\limits_{p} {_{\bot }^{n}}(p_{\neg \sigma }^{n}a_{\sigma }^{n})_{\bot }^{n})_{\bot }^{n}. \\ 
\end{array}\]

\noindent A further special case of this is where $\sigma = \bot$ too.

The next theorem is crucial to the appropriateness of the present infinite Boolean structure for a semantics of the simplified $\lambda \mu$-calculus:
\medskip

\noindent \textbf{Theorem 3} (Type Reduction Theorem) \ 
For any type $\sigma $, $D_{\neg \neg \sigma }^{n}$ is isomorphic to $D_{\sigma }^{n+1}$.\medskip

\noindent \textit{Proof.} \ $D_{\neg \sigma}^0$ is the set of functions from $D_{\sigma}^0$ to $\{0,1\}$. This may be considered $\wp (D_{\sigma}^0)$, i.e., the set of sets of the members of $D_{\sigma}^0$. Consequently, $D_{\neg \neg \sigma}^0$ may be considered $\wp (\wp (D_{\sigma}^0))$, whose members are the sets of sets of the members of $D_{\sigma}^0$. There is a one-to-one translation from those members to the member of $D_{\sigma}^1$: Read the members of each set disjunctively with $\sqcup_{\sigma}^1$, and read each of those members as the conjunction, with $\sqcap_{\sigma}^1$, of the literals, with $-_{\sigma}^1$, whose atoms are in $D_{\sigma}^0$, depending on whether or not each atom is in the original set.  For instance, if $D_{\sigma}^0 = \{a,b,c\}$, read $\{\{a\}, \{b,c\}\}$ as $(({a_{\sigma}^0}{\sqcap _{\sigma}^1}({-_{\sigma}^1}{b_{\sigma}^0})_{\sigma}^1{\sqcap _{\sigma}^1}({-_{\sigma}^1}{c_{\sigma}^0})_{\sigma}^1)_{\sigma}^1{\sqcup_{\sigma}^1}(({-_{\sigma}^1}{a_{\sigma}^0})_{\sigma}^1{\sqcap_{\sigma}^1}{b_{\sigma}^0}{\sqcap_{\sigma}^1}{c_{\sigma}^0})_{\sigma}^1)_{\sigma}^1$. That is, the resulting item is in disjunctive normal form. Since every member of $D_{\sigma}^1$ can be expressed in disjunctive normal form, there is a one-to-one correspondence between $D_{\neg \neg \sigma}^0$ and $D_{\sigma}^1$. Furthermore, the correspondence is an isomorphism because $-, \cap$, and $\cup$ in $D_{\neg \neg \sigma}^0$ are translated into $-_{\sigma}^1,  \sqcap_{\sigma}^1$, and $\sqcap_{\sigma}^1$ in $D_{\sigma}^1$. The isomorphism between $D_{\neg \neg \sigma}^n$ and $D_{\sigma}^{n+1}$, where $n>0$, is obvious. $\Box$ \medskip

This isomorphism from $D_{\neg \neg \sigma }^{n}$ to $D_{\sigma }^{n+1}$ is named $i_{\neg \neg \sigma \to \sigma}^{n+1}$. Theorem 3 and $i$ are the key to the current semantics of the simplified $\lambda \mu$-calculus. When $(\lambda x_{\neg \sigma }.P_{\bot })_{\neg \neg \sigma }$ denotes a member $a_{\neg \neg \sigma }^{n}$ of $D_{\neg \neg \sigma }^{n}$, the corresponding $(\mu _{\neg \sigma }.P_{\bot })_{\sigma }$ denotes the shadow $(i_{\neg \neg \sigma \to \sigma}^{n+1}a_{\neg \neg \sigma }^{n})_{\sigma}^{n+1}$ in $D_{\sigma }^{n+1}$. The guaranteed existence of such shadows makes double-negation elimination possible in classical type inferences. By a repeated application of Theorem 3, the following follows:\medskip

\noindent \textbf{Corollary 4} \ For any type $\sigma $, $D_{\underbrace{\neg \cdots \neg }_{2m}\sigma }^{n}$ is isomorphic to $D_{\sigma }^{n+m}$, and $D_{\underbrace{\neg \cdots \neg }_{2m+1}\sigma }^{n}$ is isomorphic to $D_{\neg \sigma }^{n+m}$.\medskip

\noindent The picture that emerges from the above series of results is as follows:\medskip

$\ \ \quad \quad \quad \quad \quad \quad \quad \ \ \vdots $

$\ \ \quad \quad \quad \quad {{D}_{\neg 5 \sigma }}\supset \cdots \supset D_{\neg 5 \sigma }^{0}\quad \quad \quad \quad \quad \quad \quad \quad \quad \quad \quad \quad \quad \quad \quad \quad \quad \quad \ \ \ \vdots $

$\ \ \quad \quad \quad \quad \quad \quad \quad \quad \quad \ \downarrow i\quad \quad \quad \quad \quad \quad \quad \quad \quad \quad \quad \quad \quad \quad \quad \ \ \ \ D_{\neg 4 \sigma }^{0}\subset \cdots \subset {{D}_{\neg 4 \sigma }}$

$\ \ \quad \quad \quad \quad {{D}_{\neg 3 \sigma }}\supset \cdots \supset D_{\neg 3 \sigma }^{1}\supset D_{\neg 3 \sigma }^{0}\quad \quad \quad \quad \quad \quad \quad \quad \quad \quad \quad \quad \ \downarrow i $

$\ \ \quad \quad \quad \quad \quad \quad \quad \quad \quad \ \downarrow i \quad \quad \downarrow i \quad \quad \quad \quad \quad \quad \quad \quad \quad \ \ \ D_{\neg \neg \sigma }^{0}\subset D_{\neg \neg \sigma }^{1}\subset \cdots \subset {{D}_{\neg \neg \sigma }} $

$\ \ \quad \quad \quad \quad {{D}_{\neg \sigma }} \ \supset \cdots \supset D_{\neg \sigma }^{2} \ \supset \ D_{\neg \sigma }^{1} \ \supset \ D_{\neg \sigma }^{0}\quad \quad \quad \quad \quad \quad \ \downarrow i\quad \quad \downarrow i $

$\ \ \quad \quad \quad \quad \quad \quad \quad \quad \quad \quad \quad \quad \quad \quad \quad \quad \quad \quad \quad \ \ \ {{D}_{\sigma }^{0} \ \ \subset \ \ D_{\sigma }^{1} \ \ \subset \ \ D_{\sigma }^{2} \ \ \subset \ \cdots \ \subset {{D}_{\sigma }}}  $\medskip
 
\noindent Here \textquoteleft $\neg 3$' = \textquoteleft $\neg \neg \neg$', etc. The members of $D_{\neg \sigma}^{0}$ distribute over the members of $D_{\sigma}^{0}$, the members of $D_{\neg \neg \sigma}^{0}$ distribute over the members of $D_{\neg \sigma}^{0}$, etc. But, by $i$, the members of $D_{\sigma}^{1}$ distribute over the members of $D_{\neg \sigma}^{0}$, and the members of  $D_{\neg \sigma}^{1}$ distribute over the members of $D_{\neg \neg \sigma}^{0}$, etc.

\subsection{Interpretation}

In what follows, $\rho$ is an assignment (or a valuation) to the variables; $\rho[x\mapsto a]$ is the same assignment as $\rho$ except that the assignment to the variable $x$ is $a$.\medskip
\begin{itemize}
\item $[\![{{x}_{\sigma }}]\!]_{\rho }^{{\mathcal{M}}}=\rho (x_{\sigma})\in D_{\sigma }.$ \medskip
\item $[\![{{({{P}_{\sigma \to \tau }}{{A}_{\sigma }})}_{\tau }}]\!]_{\rho }^{{\mathcal{M}}}=[\![{{P}_{\sigma \to \tau }}]\!]_{\rho }^{{\mathcal{M}}}[\![{{A}_{\sigma }}]\!]_{\rho }^{{\mathcal{M}}}.$ \medskip
\item $[\![{{(\lambda {{x}_{\sigma }}.{{P}_{\tau }})}_{\sigma \to \tau }}]\!]_{\rho }^{{\mathcal{M}}}=$ the function $f_{\sigma \to \tau}$ such that for any $a_{\sigma }$, ${(f_{\sigma \to \tau}a_{\sigma})}_{\tau}=[\![{{P}_{\tau }}]\!]_{\rho [x\mapsto a]}^{{\mathcal{M}}}$. \medskip
\item $[\![{{(\mu {{x}_{\neg \sigma }}.{{P}_{\bot }})}_{\sigma }}]\!]_{\rho }^{{\mathcal{M}}}=$ the compound function $i_{\neg\neg\sigma \to \sigma}\circ f_{\neg \neg \sigma }$ such that $i_{\neg\neg\sigma \to \sigma}$ is the isomorphism introduced in the last subsection and $f_{\neg \neg \sigma}$ is the function such that for any $a_{\neg \sigma }$, ${(f_{\neg \neg \sigma}a_{\neg \sigma })}_{\bot}=[\![{{P}_{\bot }}]\!]_{\rho [x\mapsto a]}^{{\mathcal{M}}}$ (i.e., $f_{\neg \neg \sigma} = [\![{{(\lambda {{x}_{\neg \sigma }}.{{P}_{\bot }})}_{\neg \neg \sigma }}]\!]_{\rho }^{{\mathcal{M}}}$).\smallskip

Equivalently, $[\![{{(\mu {{x}_{\neg \sigma }}.{{P}_{\bot }})}_{\sigma }}]\!]_{\rho }^{{\mathcal{M}}}=$ the object such that for any $a_{\neg \sigma}$, $(a_{\neg \sigma}[\![{{(\mu {{x}_{\neg \sigma }}.{{P}_{\bot }})}_{\sigma }}]\!]_{\rho }^{{\mathcal{M}}})_{\bot} \linebreak= ([\![{{(\lambda {{x}_{\neg \sigma }}.{{P}_{\bot }})}_{\neg \neg \sigma }}]\!]_{\rho }^{{\mathcal{M}}}a_{\neg \sigma})_{\bot}$. \medskip
\item $\mathcal{M}, \rho \ \vDash \ P=Q$ \ iff \ $[\![P]\!]_\rho^{\mathcal{M}} $ = $[\![Q]\!]_\rho^{\mathcal{M}} $.  \medskip
\item $\Gamma \ \vDash \ N$, where $\Gamma$ is a set of equations and $N$ is an equation, iff, for any model ${\mathcal{M}}$ and assignment $\rho$, if $\mathcal{M}, \rho \ \vDash \ M$ for every $M \in \Gamma$, then $\mathcal{M}, \rho \ \vDash \ N$.
\end{itemize}

\subsection{Soundness}
The appropriateness of the above semantics for the simplified $\lambda \mu$-calculus is revealed in the following soundness theorem:\medskip

\noindent \textbf{Theorem 5} (Soundness) \ 
The simplified $\lambda \mu$-calculus is sound (or correct) with respect to the above semantics; that is, if $\Gamma \ {\vdash_{S \lambda\mu}} \ N$, then $\Gamma \ \vDash \ N$.\medskip

\noindent \textit{Proof.} \ It is obvious that all the inference rules of the theory of $\beta \eta \mu$-equality except $\mu$-equality are correct with respect to the current semantics. This includes the correctness of 11. $\eta_{\mu}$-equality:
\begin{equation}
[\![{{(\mu {{x}_{\neg \sigma }}{{({{x}_{\neg \sigma }}{{P}_{\sigma }})}_{\bot }})}_{\sigma }}]\!]_{\rho }^{\mathcal{M}}=[\![{{P}_{\sigma }}]\!]_{\rho }^{\mathcal{M}}\quad \mathrm{if} \ {{x}_{\neg \sigma }}\notin FV({{P}_{\sigma }}).
\end{equation}
The only thing left to show is thus the correctness of 12. $\mu$-equality.

$\mu$-equality given in Subsection 2.2 can also be expressed thus:
\begin{equation}
\Rightarrow {{({{(\mu {{x}_{\neg (\sigma \to \tau )}}.{{C_{\bot}[{({{x}_{\neg (\sigma \to \tau )}}{{R}_{\sigma \to \tau }})_{\bot }}]}})}_{\sigma \to \tau }}{{Q}_{\sigma }})}_{\tau }} = {{(\mu {{y}_{\neg \tau }}.{{C_{\bot}[{{({{y}_{\neg \tau }}{({{R}_{\sigma \to \tau }}{{Q}_{\sigma }})_{\tau }})}_{\bot }}]}})}_{\tau }}
\end{equation}
where $C_{\bot}[\cdot]$ is a context: if it is filled with a term of type $\bot$, it will become a term of type $\bot$.  Then what we need to show is that for any ${\mathcal{M}}$ and $\rho$,
\begin{equation}
{[\![{{({{(\mu {{x}_{\neg (\sigma \to \tau )}}.{{C_{\bot}[{({{x}_{\neg (\sigma \to \tau )}}{{R}_{\sigma \to \tau }})_{\bot }}]}})}_{\sigma \to \tau }}{{Q}_{\sigma }})}_{\tau }}]\!]_{\rho}^{{\mathcal{M}}}}={[\![{{(\mu {{y}_{\neg \tau }}.{{C_{\bot}[{{({{y}_{\neg \tau }}{({{R}_{\sigma \to \tau }}{{Q}_{\sigma }})_{\tau }})}_{\bot }}]}})}_{\tau }} ]\!]_{\rho}^{{\mathcal{M}}}}.
\end{equation}

We assume that we are dealing with one of the largest of such substitutions; once we prove (3) for it, then the result trickles down to the smaller of such substitutions. Since both sides of (2) contain exactly the same free variables (which denote objects of the same ranks), by Proposition 1, both sides of (3), i.e., individuals or shadows, must have canonical expressions that have the same molecular structure; their difference must lie solely in their atomic structures. Thus, to show (3), it is sufficient to show that it holds for the cases in which the free variables denote individuals, i.e., objects of rank 0. 

Note that, focusing on such cases, $C_{\bot}[\cdot]$ in (3) is semantically a truth function, i.e., a function from truth values (0 or 1) to truth values. There are only four such truth functions: $T1$ (identity function): $0\mapsto 0, 1\mapsto 1$; $T2$ (negation function): $0\mapsto 1, 1\mapsto 0$; $T3$ (constant falsity function): $0,1\mapsto 0$; $T4$ (constant truth function): $0,1\mapsto 1$.

If $C_{\bot}[\cdot]=T1$, then we can simply eliminate it. Then, by (1), both the right side and the left side of the equation (3) equal ${[\![({{R}_{\sigma \to \tau }}{{Q}_{\sigma }})_{\tau }} ]\!]_{\rho}^{{\mathcal{M}}}$, and the equation holds. Similarly, if $C_{\bot}[\cdot]=T2$, both sides equal ${-_{\tau}^1}{[\![({{R}_{\sigma \to \tau }}{{Q}_{\sigma }})_{\tau }} ]\!]_{\rho}^{{\mathcal{M}}}$, and the equation holds. If $C_{\bot}[\cdot]=T3$ (or $=T4$), then, on the one hand, $[\![(\mu {y}_{\neg \tau }.C_{\bot}[({y}_{\neg \tau }({{R}_{\sigma \to \tau }}{{Q}_{\sigma }})_{\tau })_{\bot }])_{\tau } ]\!]_{\rho}^{{\mathcal{M}}}=({\bigsqcap}_{\tau}^1{a_{\tau}^0})_{\tau}^1$ (resp. $=({\bigsqcup}_{\tau}^1{a_{\tau}^0})_{\tau}^1$), i.e., the infimum (supremum) of $D_{\tau}^1$.  On the other hand, $[\![{(\mu {x}_{\neg (\sigma \to \tau )}.{C_{\bot}[({x}_{\neg (\sigma \to \tau )}{{R}_{\sigma \to \tau })_{\bot }}]})_{\sigma \to \tau }}]\!]_{\rho}^{{\mathcal{M}}}=({{\bigsqcap}_{\sigma \to \tau}^1}{p_{\sigma \to \tau}^0})_{\sigma \to \tau}^1$ (resp. $=({{\bigsqcup}_{\sigma \to \tau}^1}{p_{\sigma \to \tau}^0})_{\sigma \to \tau}^1$); so, assuming that $[\![Q_{\sigma}]\!]_{\rho}^{{\mathcal{M}}}=q_{\sigma}^{0}$, ${[\![((\mu {x}_{\neg (\sigma \to \tau )}.C_{\bot}[ ({{x}_{\neg (\sigma \to \tau )}}{R}_{\sigma \to \tau })_{\bot }])_{\sigma \to \tau }{{Q}_{\sigma }})_{\tau }]\!]_{\rho}^{{\mathcal{M}}}}=[\![(\mu {{x}_{\neg (\sigma \to \tau )}}.C_{\bot}[ ({x}_{\neg (\sigma \to \tau )}{R}_{\sigma \to \tau })_{\bot }])_{\sigma \to \tau }]\!]_{\rho}^{{\mathcal{M}}}[\![Q_{\sigma}]\!]_{\rho}^{{\mathcal{M}}}=(({{\bigsqcap}_{\sigma \to \tau}^1}{p_{\sigma \to \tau}^0})_{\sigma \to \tau}^{1}q_{\sigma}^{0})_{\tau}^1=(({{\bigsqcap}_{\tau}^1}({p_{\sigma \to \tau}^0}q_{\sigma}^{0})_{\tau}^0)_{\tau}^{1} = \linebreak({{\bigsqcap}_{\tau}^1}{a_{\tau}^0})_{\tau}^1$ (resp. $=({{\bigsqcup}_{\tau}^1}{a_{\tau}^0})_{\tau}^1$). Thus, again, the equation holds.  Therefore, regardless of what $C_{\bot}[\cdot]$ is, the equation holds. $\Box$

\section{The classical type system}
We regret to say that we have not yet determined whether or not the simplified $\lambda \mu$-calculus is complete with respect to the above semantics -- that is, whether or not if $\Gamma \ \vDash \ N$, then $\Gamma \ {\vdash_{S \lambda\mu}} \ N$. We wish to point out, however, that the infinitely nested Boolean structure of the above semantics suggests a new and interesting deductive system of classical type theory. We call it simply \textit{the classical type system (CTS)}. It is a sequent calculus. In this section we shall present \textit{CTS}.

The distinctive features of \textit{CTS} are as follows: 

\begin{enumerate}
\item Its terms, i.e., the members of each sequent, are expressions (subterms) of type $\bot$, which are identified (naturally) as propositions. 
\item Its language is purely combinatory and does not include $\lambda$, $\mu$, or bound variables. 
\item The language, instead, includes the usual logical operators, $\neg$, $\wedge$, $\vee$, $\bigwedge$, and $\bigvee$ (the universal and the existential quantifier without binding, taken as the generalized conjunction and disjunction), which simulate $-$, $\sqcap$, $\sqcup$, $\bigsqcap$, and $\bigsqcup$ in the previous semantics.
\item The deduction rules for each logical operators consist of six rules: two usual introduction rules (for the antecedent and the succeedent), and four substitution rules (for the antecedent and the succeedent) corresponding to the expansion conditions presented in Subsection 3.1.
\end{enumerate}

We now present \textit{CTS}.

\subsection{Language}
\begin{itemize}
\item Types. Same as those in $S \lambda \mu$.
\item Variables. Same as those in $S \lambda \mu$.
\item Subterms. 
\[P_{\sigma }^{k}\ ::=\ x_{\sigma }^{k}\ |\ (P_{\sigma \to \tau }^{m}P_{\sigma }^{n})_{\tau }^{\max (m,n)}\ |\ (\neg _{\sigma }^{k}P_{\sigma }^{m})_{\sigma }^{k}\ |\ (P_{\sigma }^{m}\wedge _{\sigma }^{k}P_{\sigma }^{n})_{\sigma }^{k}\ |\ (P_{\sigma }^{m}\vee _{\sigma }^{k}P_{\sigma }^{n})_{\sigma }^{k}\ |\ (\bigwedge {_{\sigma }^{k}}x_{\sigma }^{m})_{\sigma }^{k}\ |\ (\bigvee{ _{\sigma }^{k}}x_{\sigma }^{m})_{\sigma }^{k}\]
where $m,n\leq k \neq 0$.
\item Terms. 
The terms are the subterms of type $\bot$.
\end{itemize}

\subsection{Deduction system}
To avoid unnecessary repetition, in what follows only the rules for negation and conjunction are presented (along with the axiom), but the rules for the other operators are analogous. The following rules may seem to be rather complicated at first sight, but the only important point is that the rank $k$ of the operator introduced needs to be the highest in the subterm.\bigskip
\begin{center}

\AxiomC{}
\RightLabel{$Ax$}
\UnaryInfC{$\Gamma ,\ {{A}_{\bot }^{k}}\Rightarrow {{A}_{\bot }^{k}}, \, \Delta$}
\DisplayProof
\medskip

\AxiomC{$\Gamma \Rightarrow A_{\bot}^{m}, \ \Delta$}
\RightLabel{$\neg L \bot$}
\UnaryInfC{$\Gamma, \ (\neg_{\bot}^{k}A_{\bot}^{m})_{\bot}^{k} \Rightarrow \Delta$}
\DisplayProof
\quad \quad \quad \quad \quad 
\AxiomC{$\Gamma,\ A_{\bot}^{m} \Rightarrow \Delta$}
\RightLabel{$\neg R \bot$}
\UnaryInfC{$\Gamma \Rightarrow (\neg_{\bot}^{k}A_{\bot}^{m})_{\bot}^{k}, \ \Delta$}
\DisplayProof
\medskip

In the above two rules, $m \leq k \neq 0$.
\medskip

\AxiomC{$\Gamma ,\ C_{\bot}[(\neg _{\tau }^{k}(R_{\sigma \to \tau }^{h}A_{\sigma }^{m})_{\tau }^{\max (h,m)})_{\tau }^{k}]\Rightarrow \Delta $} 
\doubleLine
\RightLabel{$\neg Lr$}
\UnaryInfC{$\Gamma ,\ C_{\bot}[(R_{\sigma \to \tau }^{h}(\neg _{\sigma }^{k}A_{\sigma }^{m})_{\sigma }^{k})_{\tau }^{k}]\Rightarrow \Delta $}
\DisplayProof
\quad \quad \quad \quad \quad
\AxiomC{$\Gamma \Rightarrow C_{\bot}[(\neg _{\tau }^{k}(R_{\sigma \to \tau }^{h}A_{\sigma }^{m})_{\tau }^{\max (h,m)})_{\tau }^{k}], \ \Delta $} 
\doubleLine
\RightLabel{$\neg Rr$}
\UnaryInfC{$\Gamma \Rightarrow C_{\bot}[(R_{\sigma \to \tau }^{h}(\neg _{\sigma }^{k}A_{\sigma }^{m})_{\sigma }^{k})_{\tau }^{k}], \ \Delta $}
\DisplayProof
\medskip

In the above two rules, $h+1, m \leq k$.
\medskip

\AxiomC{$\Gamma ,\ C_{\bot}[(\neg _{\tau }^{k}(P_{\sigma \to \tau }^{m}A_{\sigma }^{h})_{\tau }^{\max (m,h)})_{\tau }^{k}]\Rightarrow \Delta $} 
\doubleLine
\RightLabel{$\neg Ll$}
\UnaryInfC{$\Gamma ,\ C_{\bot}[((\neg _{\sigma \to \tau }^{k}P_{\sigma \to \tau }^{m})_{\sigma \to \tau }^{k}A_{\sigma }^{h})_{\tau }^{k}]\Rightarrow \Delta $}
\DisplayProof
\quad \quad \quad \quad \quad
\AxiomC{$\Gamma \Rightarrow C_{\bot}[(\neg _{\tau }^{k}(P_{\sigma \to \tau }^{m}A_{\sigma }^{h})_{\tau }^{\max (m,h)})_{\tau }^{k}], \ \Delta $} 
\doubleLine
\RightLabel{$\neg Rl$}
\UnaryInfC{$\Gamma \Rightarrow C_{\bot}[((\neg _{\sigma \to \tau }^{k}P_{\sigma \to \tau }^{m})_{\sigma \to \tau }^{k}A_{\sigma }^{h})_{\tau }^{k}], \ \Delta $}
\DisplayProof
\medskip

In the above two rules $h, m \leq k \neq 0$.
\medskip

\AxiomC{$\Gamma, \ A_{\bot}^{m}, \ B_{\bot}^{n} \Rightarrow \Delta$}
\RightLabel{$\wedge L \bot$}
\UnaryInfC{$\Gamma, \ (A_{\bot}^{m}\wedge_{\bot}^{k} \ B_{\bot}^{n})_{\bot}^{k} \Rightarrow \Delta$}
\DisplayProof
\quad \quad \quad \quad \quad
\AxiomC{$\Gamma \Rightarrow A_{\bot}^{m}, \ \Delta$}
\AxiomC{$\Gamma \Rightarrow B_{\bot}^{n}, \ \Delta$}
\RightLabel{$\wedge R \bot$}
\BinaryInfC{$\Gamma \Rightarrow (A_{\bot}^{m}\wedge_{\bot}^{k}B_{\bot}^{n})_{\bot}^{k}, \ \Delta$}
\DisplayProof
\medskip

In the above two rules, $m, n \leq k \neq 0$.\medskip

\AxiomC{$\Gamma ,\ C_{\bot}[((R_{\sigma \to \tau }^{h}A_{\sigma }^{m})_{\tau }^{\max (h,m)}\wedge _{\tau}^{k}(R_{\sigma \to \tau }^{h}B_{\sigma }^{n})_{\tau }^{\max (h,n)})_{\tau }^{k}]\Rightarrow \Delta $}
\doubleLine
\RightLabel{$\wedge Lr$}
\UnaryInfC{$\Gamma ,\ C_{\bot}[(R_{\sigma \to \tau }^{h}(A_{\sigma }^{m}\wedge _{\sigma }^{k}B_{\sigma }^{n})_{\sigma }^{k})_{\tau }^{k}]\Rightarrow \Delta $}
\DisplayProof
\medskip

\AxiomC{$\Gamma \Rightarrow C_{\bot}[((R_{\sigma \to \tau }^{h}A_{\sigma }^{m})_{\tau }^{\max (h,m)}\wedge _{\tau}^{k}(R_{\sigma \to \tau }^{h}B_{\sigma }^{n})_{\tau }^{\max (h,n)})_{\tau }^{k}], \ \Delta $}
\doubleLine
\RightLabel{$\wedge Rr$}
\UnaryInfC{$\Gamma \Rightarrow C_{\bot}[(R_{\sigma \to \tau }^{h}(A_{\sigma }^{m}\wedge _{\sigma }^{k}B_{\sigma }^{n})_{\sigma }^{k})_{\tau }^{k}], \ \Delta $}
\DisplayProof
\medskip

In the above two rules, $h+1, m, n \leq k$.\medskip

\AxiomC{$\Gamma ,\ C_{\bot}[((P_{\sigma \to \tau }^{m}A_{\sigma }^{h})_{\tau }^{\max (m,h)}\wedge _{\tau}^{k}(Q_{\sigma \to \tau }^{n}A_{\sigma }^{h})_{\tau }^{\max (n,h)})_{\tau }^{k}]\Rightarrow \Delta $}
\doubleLine
\RightLabel{$\wedge Ll$}
\UnaryInfC{$\Gamma ,\ C_{\bot}[((P_{\sigma \to \tau }^{m}\wedge _{\sigma \to \tau }^{k}Q_{\sigma \to \tau }^{n})_{\sigma \to \tau }^{k}A_{\sigma }^{h})_{\tau }^{k}]\Rightarrow \Delta$
}
\DisplayProof
\medskip

\AxiomC{$\Gamma \Rightarrow C_{\bot}[((P_{\sigma \to \tau }^{m}A_{\sigma }^{h})_{\tau }^{\max (m,h)}\wedge _{\tau}^{k}(Q_{\sigma \to \tau }^{n}A_{\sigma }^{h})_{\tau }^{\max (n,h)})_{\tau }^{k}], \ \Delta $}
\doubleLine
\RightLabel{$\wedge Rl$}
\UnaryInfC{$\Gamma \Rightarrow C_{\bot}[((P_{\sigma \to \tau }^{m}\wedge _{\sigma \to \tau }^{k}Q_{\sigma \to \tau }^{n})_{\sigma \to \tau }^{k}A_{\sigma }^{h})_{\tau }^{k}], \ \Delta$
}
\DisplayProof
\medskip

In the above two rules, $h, m, n \leq k \neq 0$.

\end{center}\medskip

\noindent Here $\Gamma$ and $\Delta$ are sets of terms. $C_{\bot}[\cdot]$ is a context: if it is filled with an appropriate subterm, it will become a term. As is usually the case in sequent calculi, the antecedents of $\Rightarrow$ should be read conjunctively, and the succedents disjunctively. We regard $\Gamma \ {\vdash_{CTS}} \ \Delta$ if and only if \textquoteleft $\Gamma \Rightarrow \Delta$' is provable with the above inference rules.

\subsection{Semantics}
The domains of the models are the same as those given in Subsection 3.1, and the interpretations of subterms are straightforward as follows:\medskip

\begin{itemize}
\item $[\![x_{\sigma }^{k}]\!]_{\rho }^{{\mathcal{M}}}=\rho (x_{\sigma }^{k})\in D_{\sigma }^{k}$.\medskip
\item $[\![(P_{\sigma \to \tau }^{m}A_{\sigma }^{n})_{\tau }^{\max(m, n)}]\!]_{\rho }^{{\mathcal{M}}}=[\![P_{\sigma \to \tau }^{m}]\!]_{\rho }^{{\mathcal{M}}}[\![A_{\sigma }^{n}]\!]_{\rho }^{{\mathcal{M}}}$.\medskip
\item $[\![(\neg _{\sigma }^{k}A_{\sigma }^{m})_{\sigma }^{k}]\!]_{\rho }^{{\mathcal{M}}}=-_{\sigma }^{k}[\![A_{\sigma }^{m}]\!]_{\rho }^{{\mathcal{M}}}$.\medskip
\item $[\![(A_{\sigma }^{m}\wedge _{\sigma }^{k}B_{\sigma }^{n})_{\sigma }^{k}]\!]_{\rho }^{{\mathcal{M}}}=[\![A_{\sigma }^{m}]\!]_{\rho }^{{\mathcal{M}}}\sqcap _{\sigma }^{k}[\![B_{\sigma }^{n}]\!]_{\rho }^{{\mathcal{M}}}$\  and \ $[\![(A_{\sigma }^{m}\vee _{\sigma }^{k}B_{\sigma }^{n})_{\sigma }^{k}]\!]_{\rho }^{{\mathcal{M}}}=[\![A_{\sigma }^{m}]\!]_{\rho }^{{\mathcal{M}}}\sqcup _{\sigma }^{k}[\![B_{\sigma }^{n}]\!]_{\rho }^{{\mathcal{M}}}$.\medskip
\item $[\![(\bigwedge _{\sigma }^{k}x_{\sigma }^{m})_{\sigma }^{k}]\!]_{\rho }^{{\mathcal{M}}}=\bigsqcap_{\sigma }^{k}[\![x_{\sigma }^{m}]\!]_{\rho }^{{\mathcal{M}}}$ \ and \ $[\![(\bigvee _{\sigma }^{k}x_{\sigma }^{m})_{\sigma }^{k}]\!]_{\rho }^{{\mathcal{M}}}=\bigsqcup_{\sigma }^{k}[\![x_{\sigma }^{m}]\!]_{\rho }^{{\mathcal{M}}}$.
\end{itemize}\medskip

\textit{CTS} is clearly sound and complete with respect to this semantics, which basically states that \textit{CTS} ought to be the mirror image of the valid object-level inferences involving shadows.

\section{Note on the unranked classical type system}
Since $S \lambda \mu$ is unranked, comparing to it, it seems only natural to think of the unranked \textit{CTS}, \textit{UCTS}, dropping superscripts from the subterms and terms of \textit{CTS}.

\subsection{Language}
\begin{itemize}
\item Types. Same as those in \textit{CTS}.
\item Variables. Same as those in \textit{CTS}.
\item Subterms. 
\[{{P}_{\sigma }}::={{x}_{\sigma }} \ | \ {{({{P}_{\sigma \to \tau }}{{P}_{\sigma }})}_{\tau }} \ | \ {(\neg_{\sigma} P_{\sigma})}_{\sigma} \ | \ {(P_{\sigma} \wedge_{\sigma} P_{\sigma})}_{\sigma} \ | \ {(P_{\sigma} \vee_{\sigma} P_{\sigma})}_{\sigma} \ | \ {({\bigwedge}_{\sigma} x_{\sigma})}_{\sigma} \ | \ {({\bigvee}_{\sigma} x_{\sigma})}_{\sigma}\]
\item Terms. 
The terms are the subterms of type $\bot$.
\end{itemize}

\subsection{Deduction system}
\medskip
\begin{center}

\AxiomC{}
\RightLabel{$Ax$}
\UnaryInfC{$\Gamma ,\ {{A}_{\bot }}\Rightarrow {{A}_{\bot }}, \, \Delta$}
\DisplayProof
\medskip

\AxiomC{$\Gamma \Rightarrow A_{\bot}, \ \Delta$}
\RightLabel{$\neg L \bot$}
\UnaryInfC{$\Gamma, \ (\neg_{\bot}A_{\bot})_{\bot} \Rightarrow \Delta$}
\DisplayProof
\quad \quad \quad \quad \quad 
\AxiomC{$\Gamma,\ A_{\bot} \Rightarrow \Delta$}
\RightLabel{$\neg R \bot$}
\UnaryInfC{$\Gamma \Rightarrow (\neg_{\bot}A_{\bot})_{\bot}, \ \Delta$}
\DisplayProof
\medskip

\AxiomC{$\Gamma ,\ C_{\bot}[(\neg _{\tau }(R_{\sigma \to \tau }A_{\sigma })_{\tau })_{\tau }]\Rightarrow \Delta $} 
\doubleLine
\RightLabel{$\neg Lr$}
\UnaryInfC{$\Gamma ,\ C_{\bot}[(R_{\sigma \to \tau }(\neg _{\sigma }A_{\sigma })_{\sigma })_{\tau }]\Rightarrow \Delta $}
\DisplayProof
\quad \quad \quad \quad \quad
\AxiomC{$\Gamma \Rightarrow C_{\bot}[(\neg _{\tau }(R_{\sigma \to \tau }A_{\sigma })_{\tau })_{\tau }], \ \Delta $} 
\doubleLine
\RightLabel{$\neg Rr$}
\UnaryInfC{$\Gamma \Rightarrow C_{\bot}[(R_{\sigma \to \tau }(\neg _{\sigma }A_{\sigma })_{\sigma })_{\tau }], \ \Delta $}
\DisplayProof
\medskip

\AxiomC{$\Gamma ,\ C_{\bot}[(\neg _{\tau }(P_{\sigma \to \tau }A_{\sigma })_{\tau })_{\tau }]\Rightarrow \Delta $} 
\doubleLine
\RightLabel{$\neg Ll$}
\UnaryInfC{$\Gamma ,\ C_{\bot}[((\neg _{\sigma \to \tau }P_{\sigma \to \tau })_{\sigma \to \tau }A_{\sigma })_{\tau }]\Rightarrow \Delta $}
\DisplayProof
\quad \quad \quad \quad \quad
\AxiomC{$\Gamma \Rightarrow C_{\bot}[(\neg _{\tau }(P_{\sigma \to \tau }A_{\sigma })_{\tau })_{\tau }], \ \Delta $} 
\doubleLine
\RightLabel{$\neg Rl$}
\UnaryInfC{$\Gamma \Rightarrow C_{\bot}[((\neg _{\sigma \to \tau }P_{\sigma \to \tau })_{\sigma \to \tau }A_{\sigma })_{\tau }], \ \Delta $}
\DisplayProof
\medskip

\AxiomC{$\Gamma, \ A_{\bot}, \ B_{\bot} \Rightarrow \Delta$}
\RightLabel{$\wedge L \bot$}
\UnaryInfC{$\Gamma, \ (A_{\bot}\wedge_{\bot} \ B_{\bot})_{\bot} \Rightarrow \Delta$}
\DisplayProof
\quad \quad \quad \quad \quad
\AxiomC{$\Gamma \Rightarrow A_{\bot}, \ \Delta$}
\AxiomC{$\Gamma \Rightarrow B_{\bot}, \ \Delta$}
\RightLabel{$\wedge R \bot$}
\BinaryInfC{$\Gamma \Rightarrow (A_{\bot}\wedge_{\bot}B_{\bot})_{\bot}, \ \Delta$}
\DisplayProof
\medskip

\AxiomC{$\Gamma ,\ C_{\bot}[((R_{\sigma \to \tau }A_{\sigma })_{\tau }\wedge _{\tau}(R_{\sigma \to \tau }B_{\sigma })_{\tau })_{\tau }]\Rightarrow \Delta $}
\doubleLine
\RightLabel{$\wedge Lr$}
\UnaryInfC{$\Gamma ,\ C_{\bot}[(R_{\sigma \to \tau }(A_{\sigma }\wedge _{\sigma }B_{\sigma })_{\sigma })_{\tau }]\Rightarrow \Delta $}
\DisplayProof
\quad
\AxiomC{$\Gamma \Rightarrow C_{\bot}[((R_{\sigma \to \tau }A_{\sigma })_{\tau }\wedge _{\tau}(R_{\sigma \to \tau }B_{\sigma })_{\tau })_{\tau }], \ \Delta $}
\doubleLine
\RightLabel{$\wedge Rr$}
\UnaryInfC{$\Gamma \Rightarrow C_{\bot}[(R_{\sigma \to \tau }(A_{\sigma }\wedge _{\sigma }B_{\sigma })_{\sigma })_{\tau }], \ \Delta $}
\DisplayProof
\medskip

\AxiomC{$\Gamma ,\ C_{\bot}[((P_{\sigma \to \tau }A_{\sigma })_{\tau }\wedge _{\tau}(Q_{\sigma \to \tau }A_{\sigma })_{\tau })_{\tau }]\Rightarrow \Delta $}
\doubleLine
\RightLabel{$\wedge Ll$}
\UnaryInfC{$\Gamma ,\ C_{\bot}[((P_{\sigma \to \tau }\wedge _{\sigma \to \tau }Q_{\sigma \to \tau })_{\sigma \to \tau }A_{\sigma })_{\tau }]\Rightarrow \Delta$
}
\DisplayProof
\quad
\AxiomC{$\Gamma \Rightarrow C_{\bot}[((P_{\sigma \to \tau }A_{\sigma })_{\tau }\wedge _{\tau}(Q_{\sigma \to \tau }A_{\sigma })_{\tau })_{\tau }], \ \Delta $}
\doubleLine
\RightLabel{$\wedge Rl$}
\UnaryInfC{$\Gamma \Rightarrow C_{\bot}[((P_{\sigma \to \tau }\wedge _{\sigma \to \tau }Q_{\sigma \to \tau })_{\sigma \to \tau }A_{\sigma })_{\tau }], \ \Delta$
}
\DisplayProof
\medskip
\end{center}\medskip

The deduction system of \textit{RCTS} seems to make perfect sense on its own. We simply assign a sufficiently high rank `in our head' to the logical operator introduced as we go along.  However, a compositional semantics like those given to $S \lambda \mu$ and \textit{CTS} cannot be given to \textit{RCTS} at least in a straightforward manner, for the rank of an operator cannot be determined once and for all -- a situation somewhat similar to that of the untyped $\lambda$ calculus. Further research is called for in this connection.

\section{Conclusion}
Our investigation of the simplified $\lambda \mu$-calculus has led us to the structured domains of its models, the infinitely nested Boolean structures, and to classical type system (\textit{CTS}), which reflects the structures of the domains in a more straightforward fashion. \textit{CTS} is attractive because, though it is just as basic as the simply-typed $\lambda$-calculus or combinatory logic, it incorporates the basic classical logical operators such as classical negation, conjunction, and disjunction. This is an advantage at least in certain situations because we often find it difficult to introduce those operators into basic systems of computation. Since $S \lambda \mu$ and \textit{CTS} look very different at first sight, it is all the more interesting and important to explore the deep relations between them.

We also need to answer the outstanding question whether $S \lambda \mu$ is complete with respect to our semantics, and, if not, what more rules should be added to it. Another interesting question is how to expand our semantics to deal with the untyped $\lambda \mu$-calculus. In Scott's \cite{scott} $D_{\infty}$ model of the untyped $\lambda$-calculus, only the \textit{continuous} functions are selected as the members of a function type, i.e., the functions whose suprema $\bigsqcup$ are distributable. The domains of our models are even more restrictive since the negation (or complement) is distributable as well. Can we, perhaps, use basically the same models to deal with the untyped $\lambda \mu$-calculus?

Outside of logic and computation, classical type theory may have much use in formal linguistics. According to \textit{generalized quantifier theory} (see, e.g., Montague \cite{montague} and Barwise and Cooper \cite{barwise-cooper}), proper names and their conjoinments such as \textit{Adam}, \textit{Bob}, \textit{Adam and Bob}, and \textit{Adam or Bob}, as well as quantificational phrases such as \textit{every man} and \textit{some women}, are not of the type of individuals $e$ (or $ind$ or $i$) but of type $\neg \neg e$. This is based on the idea that we cannot have logically compound objects in $e$. This theory has had much success in some areas, but has difficulty dealing with the scopes of logical operators. For instance, \medskip

\begin{itemize}
\item Adam and Bob love Carol or Diane
\end{itemize}\medskip

\noindent is ambiguous and can be read in two ways, but, assuming that the grammatical structure of the sentence is fixed as $[[\mathit{Adam \ and \ Bob}]_{NP}[[\mathit{love}]_{V}[\mathit{Carol \ and \ Diane}]_{NP}]_{VP}]_{S}$, how can we analyze the ambiguity?  The advocates of generalized quantifier theory do have a few answers, but they are all rather complicated. For instance, in Hendriks' \cite{hendriks} answer, \textit{x loves y} is given two formalizations, $\lambda {v}\lambda {z}({v}(\lambda {x}({z}(\lambda {y}.(l{y}){{x}}))))$ and $\lambda {z}\lambda {v}({v}(\lambda {y}({z}(\lambda {x}.(l{y}){x}))))$. In contrast, \textit{CTS} can offer a very simple answer:\medskip 

\begin{itemize}
\item $((L_{e\to \neg e}^{0}(C_{e}^{0}\vee _{e}^{1}D_{e}^{0})_{e}^{1})_{\neg e}^{1}(A_{e}^{0}\wedge _{e}^{1}B_{e}^{0})_{e}^{1})_{\bot }^{1}$ 
$\Rightarrow \\ ((((L_{e\to \neg e}^{0}C_{e}^{0})_{\neg e}^{0}A_{e}^{0})_{\bot }^{0}\wedge _{\bot }^{1}((L_{e\to \neg e}^{0}C_{e}^{0})_{\neg e}^{0}B_{e}^{0})_{\bot }^{0})_{\bot }^{1}\vee _{\bot }^{1}(((L_{e\to \neg e}^{0}D_{e}^{0})_{\neg e}^{0}A_{e}^{0})_{\bot }^{0}\wedge _{\bot }^{1}((L_{e\to \neg e}^{0}D_{e}^{0})_{\neg e}^{0}B_{e}^{0})_{\bot }^{0})_{\bot }^{1})_{\bot }^{1}$;\\ Adam and Bob both love Carol, or they both love Diane. \medskip
\item $((L_{e\to \neg e}^{0}(C_{e}^{0}\vee _{e}^{1}D_{e}^{0})_{e}^{1})_{\neg e}^{1}(A_{e}^{0}\wedge _{e}^{2}B_{e}^{0})_{e}^{2})_{\bot }^{2}$ 
$\Rightarrow \\ ((((L_{e\to \neg e}^{0}C_{e}^{0})_{\neg e}^{0}A_{e}^{0})_{\bot }^{0}\vee _{\bot }^{1}((L_{e\to \neg e}^{0}D_{e}^{0})_{\neg e}^{0}A_{e}^{0})_{\bot }^{0})_{\bot }^{1}\wedge _{\bot }^{2}(((L_{e\to \neg e}^{0}C_{e}^{0})_{\neg e}^{0}B_{e}^{0})_{\bot }^{0}\vee _{\bot }^{1}((L_{e\to \neg e}^{0}D_{e}^{0})_{\neg e}^{0}B_{e}^{0})_{\bot }^{0})_{\bot }^{1})_{\bot }^{2}$; \\ Adam loves either Carol or Diane, and so does Bob.\medskip
\end{itemize}

\noindent The two formalizations (before $\Rightarrow$) have the identical logical structure which matches the assumed grammatical structure; the only difference is the ranks of $\wedge$ (1 versus 2). But this difference leads to the two different interpretations (after $\Rightarrow$) of the original sentence. This is possible only because in \textit{CTS} the distribution of the operators is determined not by the type structures of the sentences but purely by the ranks involved. Generally, \textit{CTS} can separate the scope relations in sentences from the sentences' grammatical and logical structures (for more on this, see Akiba \cite{akiba}). A further investigation of \textit{CTS} is important also in this connection.

\bibliographystyle{eptcs}
\providecommand{\urlalt}[2]{\href{#1}{#2}}
\providecommand{\doi}[1]{doi:\urlalt{http://dx.doi.org/#1}{#1}}
\bibliography{generic}

\end{document}